\documentclass[prc,preprint,showpacs,tightenlines,floatfix,amsfonts,nofootinbib]{revtex4}
\usepackage{amsmath,amssymb}
\usepackage{epsfig}
\usepackage{graphicx}
\usepackage{amsmath}
\usepackage{amsfonts}
\usepackage{epstopdf}
\def\slashchar#1{\setbox0=\hbox{$#1$}     		
   \dimen0=\wd0                                 	
   \setbox1=\hbox{/} \dimen1=\wd1               	
   \ifdim\dimen0>\dimen1                        	
      \rlap{\hbox to \dimen0{\hfil/\hfil}}      	
      #1                                        	
   \else                                        	
      \rlap{\hbox to \dimen1{\hfil$#1$\hfil}}   	
      /                                         	
   \fi}

\renewcommand{\vec}{\boldsymbol}
\newcommand{\be}{\begin{equation}}
\newcommand{\ee}{\end{equation}}
\newcommand{\bear}{\begin{eqnarray}}
\newcommand{\eear}{\end{eqnarray}}
\newcommand{\ba}{\begin{array}}
\newcommand{\ea}{\end{array}}

\newcommand{\rhoc}{\rho_{\mathrm{c}}}
\newcommand{\rhop}{\rho_{\mathrm{p}}}

\newcommand{\psih}[1]{\psi^{J}_{#1}}

\begin{document}

\title{Hard Probe of Geometry and Fluctuations in Heavy Ion Collisions at $\sqrt{s_{_{NN}}}=$ 0.2, 2.76, and 5.5 TeV }

\author{Xilin Zhang} \email{zhangx4@ohio.edu}
\affiliation{Physics Department and Center for Exploration of Energy and Matter,
Indiana University, 2401 N Milo B. Sampson Lane, Bloomington, IN 47408, USA.}

\author{Jinfeng Liao} \email{liaoji@indiana.edu}
\affiliation{Physics Department and Center for Exploration of Energy and Matter,
Indiana University, 2401 N Milo B. Sampson Lane, Bloomington, IN 47408, USA.}
\affiliation{RIKEN BNL Research Center, Bldg. 510A, Brookhaven National Laboratory, Upton, NY 11973, USA.}


\begin{abstract}

\begin{description}

\item[Background:] A significant quenching of high energy jets was observed in the heavy ion collisions at the Relativistic Heavy Ion Collider (RHIC) facility, and is now confirmed at the Large Hadron Collider (LHC) facility. The RHIC plus LHC era provides a unique opportunity to study the jet-medium interaction that leads to the jet quenching, and the medium itself at different collision energies (medium temperatures). 
 
\item[Purpose:] We study the azimuthal anisotropy of jet quenching, to seek constraints on different models featuring distinct path-length and density dependences for jet energy loss, and to gain a better understanding of the medium.  

\item[Methods:] The models are fixed by using the RHIC data, and then applied to study the LHC case. A set of harmonic (Fourier) coefficients $v_{n}$ are extracted from the jet azimuthal anisotropy on a event-by-event basis.

\item[Results:] The second harmonics $v_{2}$, mostly driven by the medium's geometry, can be used to differentiate jet quenching models. Other harmonics are also compared with the LHC (2.76 TeV) data. The predictions for future LHC (5.5 TeV) run are presented.   

\item[Conclusions:] We find that a too strong path-length dependence (e.g., cubic) is ruled out by the LHC $v_{2}$ data,  while the model with a strong near-$T_c$ enhancement for the jet-medium interaction describes the data very well. It is worth pointing out that the latter model expects a less color-opaque medium at LHC.

\end{description}
\end{abstract}

\maketitle

\section{Introduction}

\subsection{Motivation}

Searching for new forms of matter is a fundamental quest. In the Standard Model's strong interaction sector described by the Quantum Chromodynamics (QCD), various forms of QCD matter may exist in nature, e.g., inside the compact stars and in the early Universe a few microseconds after the ``Big Bang''. Studying the ``condensed matter physics of QCD''  has been essential for advancing our understanding of matter and the QCD dynamics as well. A highly nontrivial prediction based on the salient feature of QCD,  the asymptotic freedom \cite{Gross:1973id}, is that we shall expect a new, deconfined, and weakly coupled phase of QCD matter at asymptotically high temperature\cite{Collins:1974ky,Lee:1974kn}. A deconfinement transition at certain temperature $T_c$  and a quark-gluon plasma (QGP) phase above  $T_c$ \cite{Shuryak:1978ij} are expected and have been extensively studied using lattice QCD simulations. Experimentally, such hot QCD matter has been created in the heavy ion collisions (``Little Bang''), with its many properties measured in the past decade at RHIC \cite{rhic_white_paper} and now also at LHC \cite{Muller:2012zq}.

The era of RHIC plus LHC provides unique opportunities for uncovering the underlying structures of the hot deconfined QCD matter and deepening our understanding of how QCD operates in a strongly interacting many-body setting. Various major findings at RHIC (with AuAu collisions up to $\sqrt{s}=200\rm GeV$) coherently hint at a strongly coupled QGP (sQGP) \cite{Gyulassy:2004zy} above but close to $T_c$. This is drastically different from the naively expected ``asymptotically free matter'' (AFM). While a full microscopic picture of the sQGP is yet to come, there have been important progresses suggesting that the QCD plasma near $T_c$ is an emergent matter dominated by dense and light (chromo-)magnetic monopoles \cite{Liao:2006ry}. These thermal monopoles are peculiar to the region above but close to $T_c$, and their condensation at $T_c$ marks the onset of confinement.  With LHC colliding PbPb at $\sqrt{s}=2.76\ \mathrm{TeV}$ now (and $\sqrt{s}=5.5\ \mathrm{TeV}$ in the future), and thus creating an even hotter QGP with much higher density, it is tempting to ask: if the sQGP at RHIC bears peculiar near-$T_c$ nature, is that quickly turned off at LHC with noticeable changes in the QGP properties (the QGP at LHC is further away from $T_c$), and how much closer is the QGP at LHC to the AFM? This paper attempts to extract insights into these questions by studying the hard probe of the medium geometry and its fluctuations at both collision energies. 

\subsection{Jet quenching and geometric tomography}

Just like the X-ray imaging of normal materials, the highly energetic partons produced in the initial binary hard collisions provide natural imaging tools to study the QGP matter. Such a partonic jet, carrying an energy much higher than the medium energy scale, could experience multiple collisions with medium constituents and lose its energy significantly (i.e., jet quenching) \cite{Wang:1991xy}. As a result the hadron production from the jet will differ from the case without medium effects (e.g., in the  proton-proton collision at the same energy). Measuring such difference is a  useful way of learning about the medium's properties  and the jet-medium interactions. A conventional observable to quantify the jet quenching is the nuclear modification factor $R_{AA}$ defined as: 
\begin{equation}\label{eqn_raa}
R_{AA}(p_t,\phi,\eta) \equiv \frac{d^2 N^{AA}/dp_td\phi d\eta}{T_{AA}\cdot d^2
\sigma^{NN}/dp_t d\phi d\eta } \ , 
\end{equation}
where in the denominator the nuclear overlap function $T_{AA}$ scales up single
Nucleon-Nucleon(NN) cross section to the Nucleus-Nucleus(AA) one according to the expected binary NN collision number. Thus a value of $R_{AA}$ smaller (larger) than unity means
suppression (enhancement) due to medium effect. For the jet quenching physics, we focus on the $R_{AA}$ measured for detected hadrons with large enough transverse momenta (e.g. $p_t>6 \rm GeV$ for RHIC and $p_t>8\rm GeV$ for LHC) and for $\eta$ at mid-rapidity (as per most detectors). The $\phi$ is the azimuthal angle of the measured hadron's transverse momentum. One may examine the $R_{AA}$ either integrated or differential in $\phi$. 

A significant suppression of high-$p_t$ hadron production was first observed 
at RHIC, with $R_{AA}$ reaching about $0.18$ in the most central collisions. Measurements of charged particle, identified hadron, heavy flavor production, photon production, triggered di-hadron correlations have coherently pointed to a medium that is extremely opaque to  ``colored'' hard probe.  For reviews see e.g. \cite{Gyulassy:2003mc}. New and extensive LHC data on various hard probe observables have shown similar strong jet quenching \cite{LHC_suppression,CMS:2012aa}, and the apt question is  whether the hotter medium becomes less opaque or not (in the context that medium density roughly doubles). 

A very powerful idea in the jet quenching study is the so-called geometric tomography \cite{Gyulassy:2000gk,Wang:2000fq}. The created hot medium is generally anisotropic in the transverse plane (perpendicular to the collision beam axis), and therefore high energy partons traversing the medium along different azimuthal directions will ``see'' different medium thickness and thus lose different amount of energy.  That will lead to a measurable anisotropy in $R_{AA}(\phi)$. It is well-known that the dominant  geometric anisotropy is the elliptic component $\sim \cos 2(\phi-\Psi_2)$ (with $\Psi_2$ related to matter anisotropy axis) and the coefficient is the $v_2$ for high $p_t$ hadrons. A lot of studies \cite{Shuryak:2001me,Liao:2008dk,Jia:2010ee,Jia:2011pi,Betz:2011tu,Liao:2011kr,Bass:2008rv,Rodriguez:2010di} have shown that such geometric observable is very sensitive to the underlying jet energy loss dynamics e.g., its dependence on the in-medium path length. Comprehensive studies in \cite{Renk:2011aa} have also shown its highly constraining power, and used it to rule out a wide class of models, leaving only a few cases with a certain hydrodynamical background that may be consistent with data. More recently the idea has been extended to systematically quantify the jet response to the geometrical anisotropy arising from the strong  fluctuations in the initial condition of heavy ion collisions \cite{Zhang:2012mi,Zhang:2012ha,Jia:2012ez}.

\subsection{Near-$T_c$ enhancement of jet quenching}

A number of studies on jet quenching azimuthal anisotropy at RHIC (specified by high-$p_t$ $v_2$), however, have shown a clear discrepancy between various model results and the PHENIX data with $p_t$ extending to $\sim 20\rm GeV$ \cite{Adler:2006bw,Shuryak:2001me} till around 2008. Previous jet quenching models, with either linear or the LPM-induced quadratic~\cite{Baier:1996sk} path-length dependence, under-predicted the high-$p_t$ $v_2$ by a significant amount (often by a factor of 2), although they can reproduce the overall opaqueness in terms of $\phi$-integrated $R_{AA}$. Efforts toward reconciling $R_{AA}$ and $v_2$ at high $p_t$ fostered a more radical proposal in Ref.~\cite{Liao:2008dk}, which breaks the assumption (taken for granted in all previous considerations) that the energy loss is simply proportional to plasma constituent density (e.g. as per entropy density $s$). 
Instead, the key insight of Ref.~\cite{Liao:2008dk} is that the jet-medium interaction has a non-trivial dependence on matter density and particularly is strongly enhanced in the near-$T_c$ region. Such enhancement, in analogy with the well-known ``critical opalescence'', is well motivated by the aforementioned emergence of magnetic monopoles in the same region \cite{Liao:2006ry}.   Phenomenologically this model for the first time achieved a simultaneous description of $R_{AA}$ and $v_2$ at high $p_t$. It should be emphasized that the effect of the near-$T_c$ enhancement on the jet quenching anisotropy is robust and generic: similar successes were reported by incorporating this enhancement in a variety of different approaches for jet quenching \cite{Renk:2010qx,Betz:2012fy,Buzzatti:2012dy}. 

For a more concrete discussion on the path-length dependence and matter-density dependence of jet quenching, let us adopt here the geometric models that have been widely used and successful in describing the gross features of jet quenching \cite{Shuryak:2001me,Liao:2008dk,Jia:2010ee,Jia:2011pi,Betz:2011tu,Liao:2011kr,Zhang:2012ha}. In such a model one assumes that the final energy $E_f$ of a jet with initial energy $E_i$ after traveling an in-medium path $\vec{P}$ (specified by the jet initial spot and momentum direction) can be parameterized as $E_f = E_i \times f_{\vec P}$ with the   $f_{\vec P}$ given by
\begin{equation} \label{Eq_fP}
f_{\vec P} = \exp\left\{ - \int_{\vec P}\, \kappa[s(l)]\, s(l)\, l^m dl  \right\} \ . 
\end{equation}
Here $s(l)$ is the local entropy density along the jet path, while the $\kappa(s)$ represents the local jet-medium interaction strength which as a property of underlying matter depends on the local density $s(l)$. We choose to explicitly separate out the density $s$ itself, and the combination $\kappa(s)\, s$ approximately corresponds to $\hat{q}$ in many jet quenching models. Different choices of $m$ and $\kappa(s)$ mean different path-length and matter-density dependences, and here we consider three classes of models. The near-$T_c$ enhancement (NTcE) model as in Refs.~\cite{Liao:2008dk,Liao:2011kr} is implemented by assuming $m=1$ (i.e. quadratic) and introducing a strong jet quenching component in the vicinity of $T_c$ (with density $s_c$ and span of $s_w$) via
\begin{eqnarray} \label{Eq_kappa}
\kappa(s)=\kappa_0 [1+ \xi\, \exp(-(s-s_c)^2/s_w^2)] \ , 
\end{eqnarray}
with $\xi=6$, $s_c=7/fm^3$, and $s_w=2/fm^3$. (see Refs.~\cite{Liao:2008dk,Liao:2011kr,Zhang:2012ha} for the details.) For contrast, we also consider two other classes of models that both assume $\kappa(s)=\kappa_0$ being a constant independent of density, while have $m=1$ and $m=2$ respectively, referred to as $\mathrm{L}^{2}$ and $\mathrm{L}^{3}$ models hereafter. The parameter $\kappa_0$ controls the overall opaqueness in each of the three models and will be fixed by $R_{AA}\approx 0.18$ in the $0-5\%$ collisions at RHIC $\sqrt{s}=200\rm GeV$. The fraction energy loss formula employed here gives a $p_{t}$ independent $R_{AA}$. Fig.~\ref{Raa_comparison} shows our $R_{AA}$ results for both RHIC ($0.2$ TeV) 
and LHC ($2.76$ TeV). The data are from both RHIC \cite{Adare:2008qa} and LHC \cite{Abelev:2012hxa,CMS:2012aa}. See the detailed discussion about this in Sec.~\ref{sec:hp}. The $\mathrm{L}^2$ model represents the generic feature of most radiative energy loss models with LPM effect, while the $\mathrm{L}^3$ model is 
motivated by certain energy loss calculations for strongly coupled Yang-Mills plasma based on the AdS/CFT correspondence \cite{Marquet:2009eq,Jia:2010ee}. It is worth mentioning, though, that at RHIC it turns out the $\mathrm{L}^3$ model is also able to describe $v_2$ at  high $p_t$ \cite{Jia:2010ee,Adare:2010sp} (also see Fig.~\ref{V2_comparison}). The reason may be that the $\mathrm{L}^3$ dependence effectively enhances the later time quenching which mimics the similar effect from the near-$T_c$ enhancement. Therefore further discrimination between the two models is needed and LHC test is crucial. 

The rest of the paper is organized as the follows. The next section will cover first a description of our event-by-event analysis, and then our major results on high-$p_t$ anisotropy for both RHIC and LHC energies. Comparisons with available data will also be provided there. For the details of our calculations and results, please see Ref.~\cite{Zhang:2012ha}. This report is then ended with a short summary.     

\section{Hard probe of geometry and fluctuations from RHIC to LHC} \label{sec:hp}
Here we report the first event-by-event quantification of the azimuthal anisotropy in jet quenching due to both geometry and fluctuations at both RHIC and LHC energies. In particular we will show results for $v_2$ at high $p_t$ from all three geometric models including the NTcE, $\mathrm{L}^{2}$ and $\mathrm{L}^{3}$ models, and compare them with data. 

 In general for a given event, there are strong initial fluctuations in both the participant density profile (which dominates the soft matter density distribution) and the collision density profile (which dominates the jet spot distribution). The jet quenching anisotropy comes from the convolution of both, and the resulting $R_{AA}(\phi)$ can be evaluated on an event-by-event basis as follows:
\begin{equation}
R_{AA} (\phi) =  <\, (f_{\vec P_{\phi}})^{n-2}  \,>_{\vec P_{\phi}} \ ,  \label{Eq_RAA}
\end{equation}
where $< \, \, >_{\vec{P}_\phi}$ means averaging over all jet paths with azimuthal orientation $\phi$ and including all the possible initial jet production spots (distributed according to the binary collision density in the same event). We notice that there are fluctuations of energy loss even for a given path $\vec P_{\phi}$ as demonstrated in e.g.~\cite{Renk:2006pk}: such fluctuations may likely weaken the jet response to geometric anisotropy and will be studied in the future. The exponent $n$ comes from reference p-p spectrum at the same collision energy: $n\approx 8.1, \ 6.0, \ \text{and} \ 5.4 $ for $\sqrt{s}=0.2, \ 2.76, \ \text{and} \ 5.5 $ TeV (see e.g. \cite{Adler:2006bw, LHC_suppression, Francois} ). 
The so-obtained $R_{AA}(\phi)$ in each event can be further Fourier decomposed  as:  
\begin{eqnarray}
R_{AA}(\phi)=R_{AA}\left(1+2\sum_{n=1}^{\infty}  v_{n}\cos[n(\phi-\psih{n})]\right) \ .
\end{eqnarray} 
The overall quenching $R_{AA}$ as well as the azimuthal harmonics $v_n$ and the corresponding n-axis $\psih{n}$ can then be determined from the above formula in each event, followed by average over events (about $10^4$ events for each impact parameter in this study). 
The second harmonic $v_2$ is the most robust and reflects the hard probe of the anisotropic geometry, while other (odd) harmonics are strongly affected by the initial state fluctuations and provide further insights into the initial conditions in addition to what have been learned from the bulk collective expansion dynamics \cite{Bulk_harmonics}. 

To perform a event-by-event simulation, we have used the standard Monte-Carlo Glauber model to generate fluctuating initial conditions \cite{miller2007} and followed most hydrodynamics literature to set the relevant implementation procedures and parameters \cite{HeinzMoreland2011,Hirano2009}. The NN cross section $\sigma_{inel}$ is set as $\sigma_{inel}=42, \ 62, \ \text{and} \ 66\rm \, mb$  for  $\sqrt{s}=0.2, \ 2.76, \ \text{and} \ 5.5\rm \, TeV$. With calculated participant density $\rhop(\vec{r}^{\perp})$ and binary collision density $\rhoc(\vec{r}^{\perp})$ in the transverse plane, the initial entropy density at equilibrium time $\tau_{0}\equiv0.6 \ \mathrm{fm}/\mathrm{c}$ (assumed the same for different energies) is scaled with $\rhop(\vec{r}^{\perp})$ at RHIC, and $(1-\delta)/2\times\rhop(\vec{r}^{\perp})+\delta\times\rhoc(\vec{r}^{\perp})$ at LHC ($\delta=0.118$ for both $2.76$ and $5.5$ TeV cases). The proportionality constants are determined from the multiplicities at different energies \cite{HeinzMoreland2011}. Energy loss in the pre-equilibrium stage is quite possible but quantitatively uncertain, and may be improved in the future with developing understanding of the thermalization process \cite{Blaizot:2011xf}. We've adopted the strategy in \cite{Jia:2010ee} to increase the pre-equilibrium density linearly in $\tau$ for $\tau<\tau_0$, and then to decrease the entropy density as $1/\tau$ for $\tau\geq \tau_0$.  The entropy density's $1/\tau$ evolution takes into account the boost-invariant longitudinal expansion of the medium. We note that for a more realistic hydrodynamic background \cite{Renk:2006sx,Renk:2011qi} the eccentricities (particularly the higher harmonics) decrease with time due to transverse expansion thus reducing the anisotropy in the jet response.

\begin{figure}
	\begin{center}
		\includegraphics[width=7.5cm,angle=0]{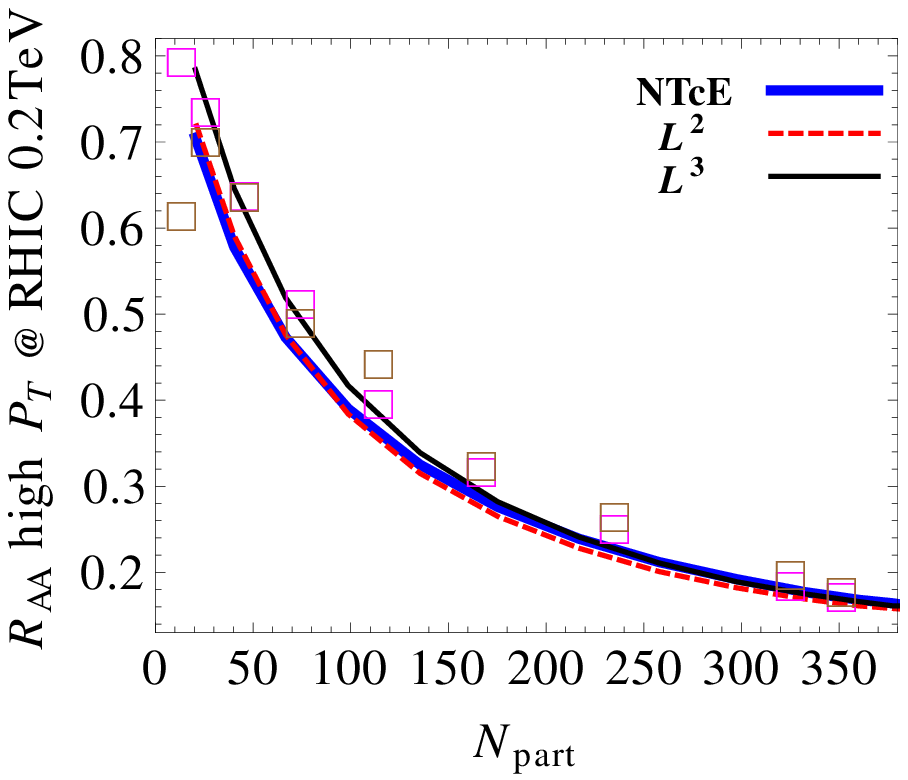} \hspace{0.5cm}
		\includegraphics[width=7.5cm,angle=0]{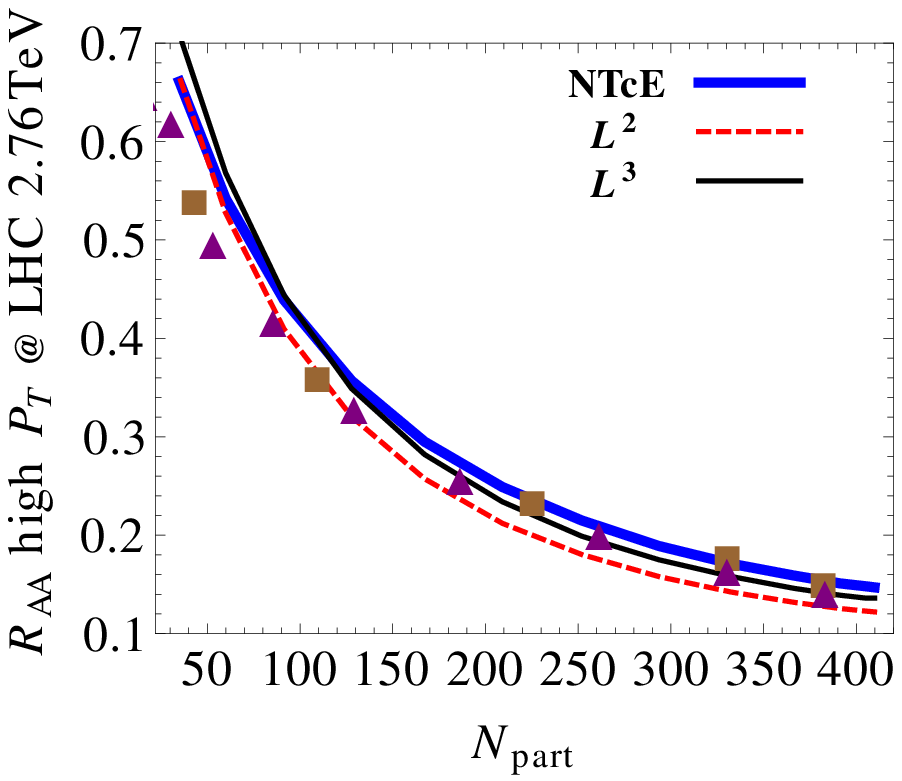}
		\caption{(color online) The $R_{AA}$ of high $p_t$ hadrons at RHIC $0.2\rm\, TeV$ (left) and LHC $2.76\rm\, TeV$ (right) collisions computed from the NTcE (blue thick solid curve),  $\mathrm{L}^{2}$ (red dashed curve) as well as $\mathrm{L}^{3}$ (black think solid curve) models. The results are compared with various data: PHENIX integrated $R_{AA}$ over $p_t>5$ GeV and $p_t>10$ GeV (magenta and brown open squares) for  the RHIC (left);  ALICE $R_{AA}$ around $8\, \rm GeV$ bin (filled purple triangles) and CMS around $8.4\, \rm GeV$ bin (filled brown box) for the LHC (right).}
		\label{Raa_comparison}
	\end{center}
	\vspace{0.cm}
\end{figure}

\begin{figure}
	\begin{center}
		\includegraphics[width=7.5cm]{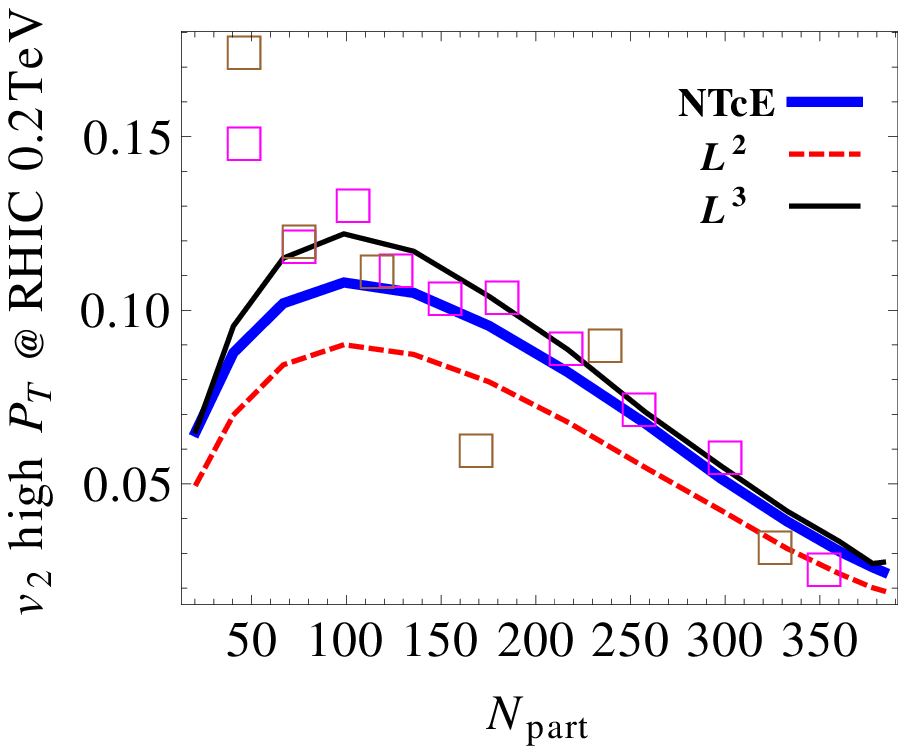}
		\includegraphics[width=7.5cm]{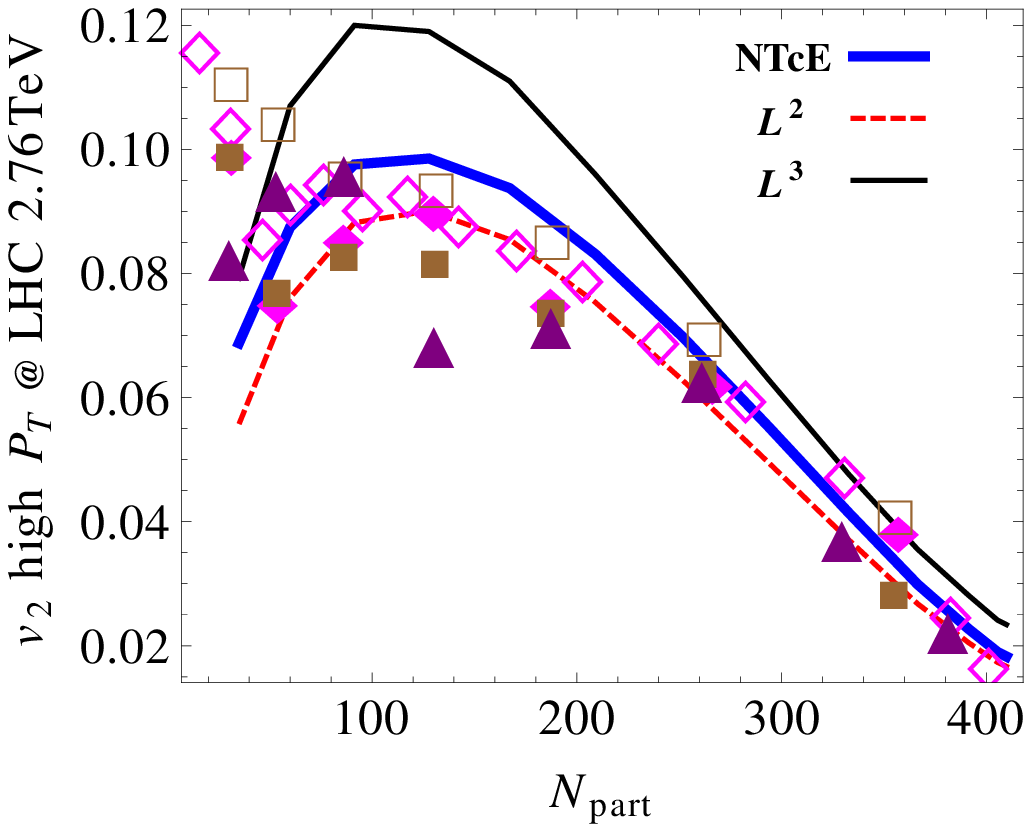}
		\caption{(color online) The $v_2$ of high $p_t$ hadrons at RHIC $0.2\rm\, TeV$ (left) and LHC $2.76\rm\, TeV$ (right) collisions computed from the NTcE (thick solid blue),  $\mathrm{L}^{2}$ (dashed red) as well as $\mathrm{L}^{3}$ (thin solid black) models. The results are compared with various data: PHENIX $6-9$ (magenta boxes) and $>9\rm \, GeV$ (brown boxes) for  the RHIC (left);  ALICE $10-20\rm \, GeV$ (purple  triangles), ATLAS $8-12$ GeV and $9-20\rm \, GeV$ (open and filled magenta diamonds), as well as CMS $9$ and $11\rm \, GeV$ bins (open and filled brown boxes) for the LHC (right).}
		\label{V2_comparison}
	\end{center}
	\vspace{-0.8cm}
\end{figure}

A first interesting observable to study is the overall $R_{AA}$: while all models are calibrated by $R_{AA}$ in the most central collisions at RHIC, how the $R_{AA}$ evolves with changing centrality and beam energy provides an important test on various models. In Fig.~\ref{Raa_comparison} we compare the high-$p_t$ $R_{AA}$ produced by the three models  with data from both RHIC \cite{Adare:2008qa} and LHC \cite{Abelev:2012hxa,CMS:2012aa}. As it turns out, however, the centrality dependence is not a very sensitive observable. At RHIC energy, all three models describe the $R_{AA}$'s centrality trend fairly well. Applied to the LHC case with increased matter density, the NTcE model predicts a nontrivial reduction of the (average) medium opaqueness, resulting in the least suppression (i.e. larger $R_{AA}$, or more ``transparency'') among the three models from mid-central to most-central  LHC events.  This reduction is due to the strong decrease of $\kappa[s]$ at higher density away from the enhancement region around $T_c$ (as is evident from Eq.(\ref{Eq_kappa}) --- see a detailed quantification of such reduction in \cite{Zhang:2012ha}). 
Such a difference is visible albeit smallish. With the present accuracy of data and modeling, it may be difficult to draw firm conclusions solely upon it. This situation thus necessitates the use of more sensitive observables to distinguish these models, which are the geometric features of jet quenching according to our study below.  

Fig.~\ref{V2_comparison} shows high-$p_t$ $v_2$  predicted by the three models and the data from both RHIC \cite{Adare:2010sp} and LHC \cite{Abelev:2012di,ATLAS:2012at,Chatrchyan:2012xq}. Most of these data bear very small statistical errors, and detailed information about statistic and systematic errors can be found in those experimental papers~\cite{Adare:2010sp,Abelev:2012di,ATLAS:2012at,Chatrchyan:2012xq}. It shall be emphasized again that all models have their parameters fixed  with RHIC $0-5\%$ data once and for all. For the RHIC comparison, the simple $\mathrm{L}^{2}$ model clearly fails to describe data while the NTcE and  $\mathrm{L}^{3}$ models are compatible with data. Moving to the LHC comparison, however, one immediately sees that too strong 
 a path-length dependence (i.e. the $\mathrm{L}^{3}$ model) is ruled out,  while both the NTcE and $\mathrm{L}^{2}$ models agree with data. Therefore, we conclude that only the model with LPM-type quadratic path-length dependence  and strong near-$T_c$ enhancement of jet-medium interaction fully describes data at both collision energies.

\begin{figure*}[t]
	\begin{center}
		\includegraphics[width=8cm]{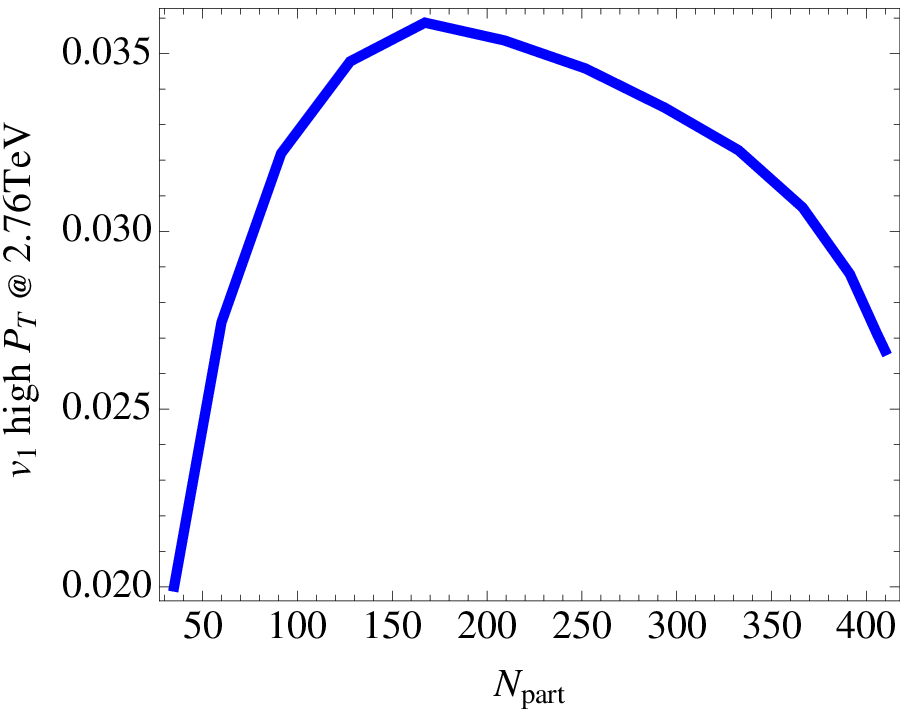}
			\includegraphics[width=8cm]{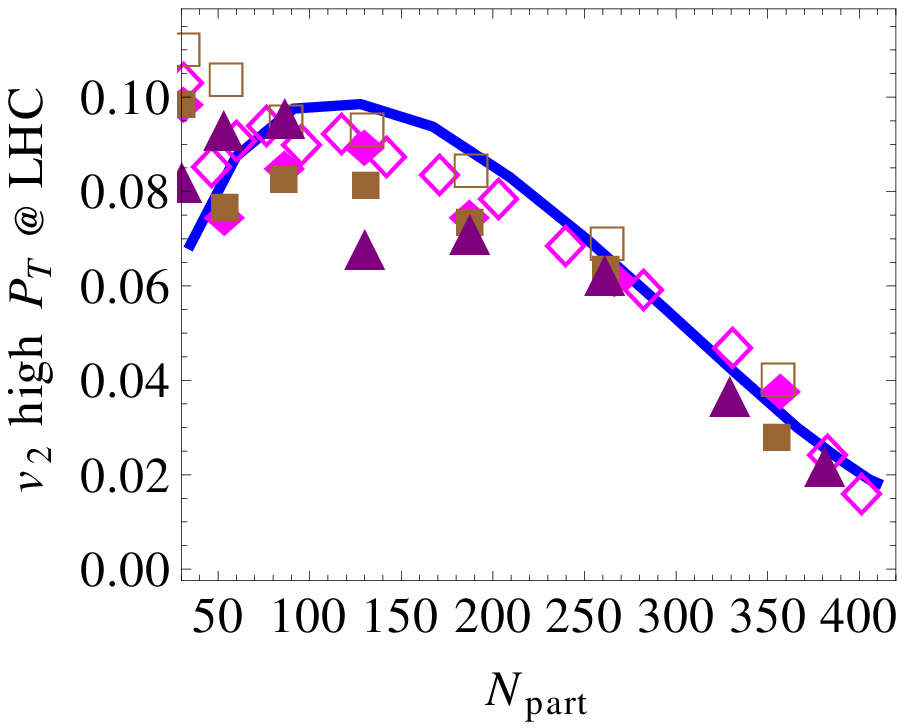}
			\includegraphics[width=8cm]{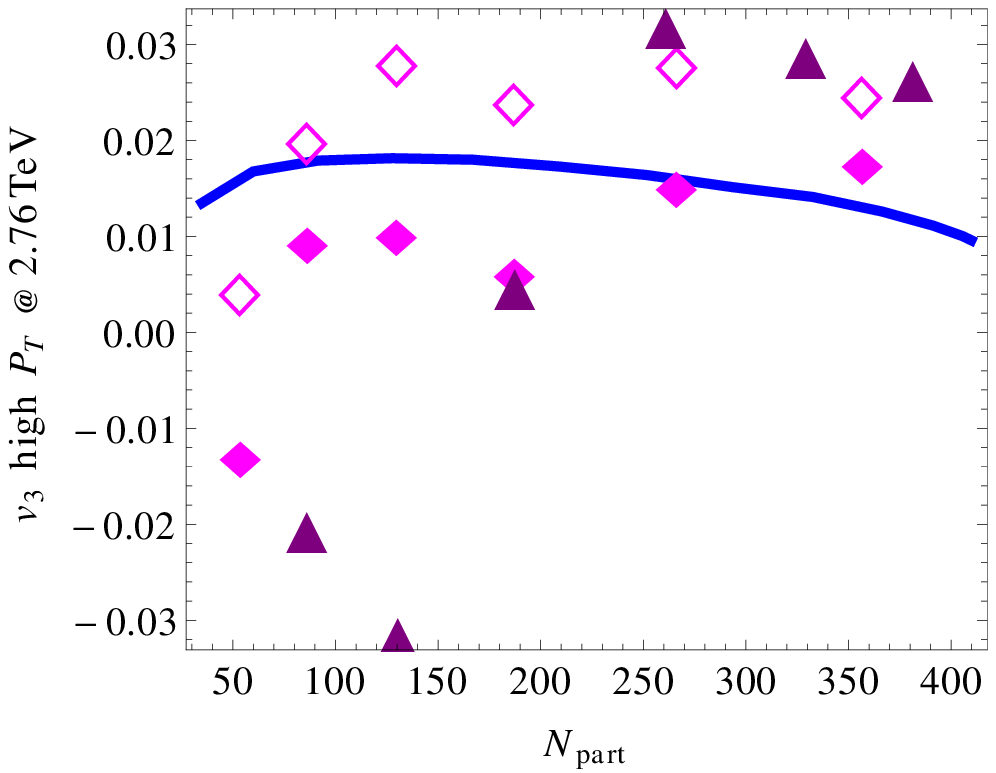}
			\includegraphics[width=8.1cm]{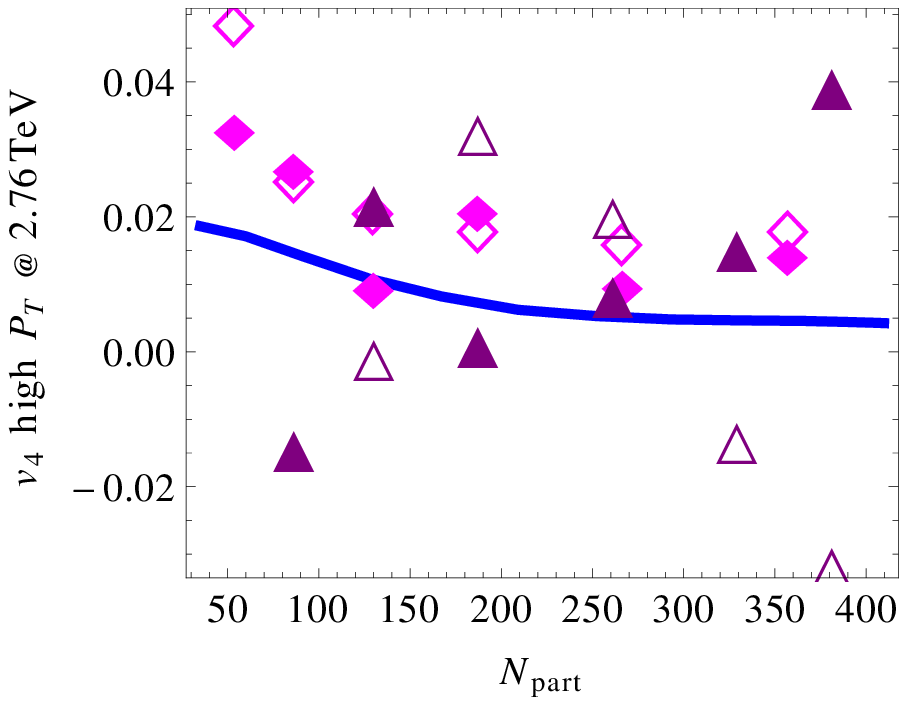}
			\includegraphics[width=8cm]{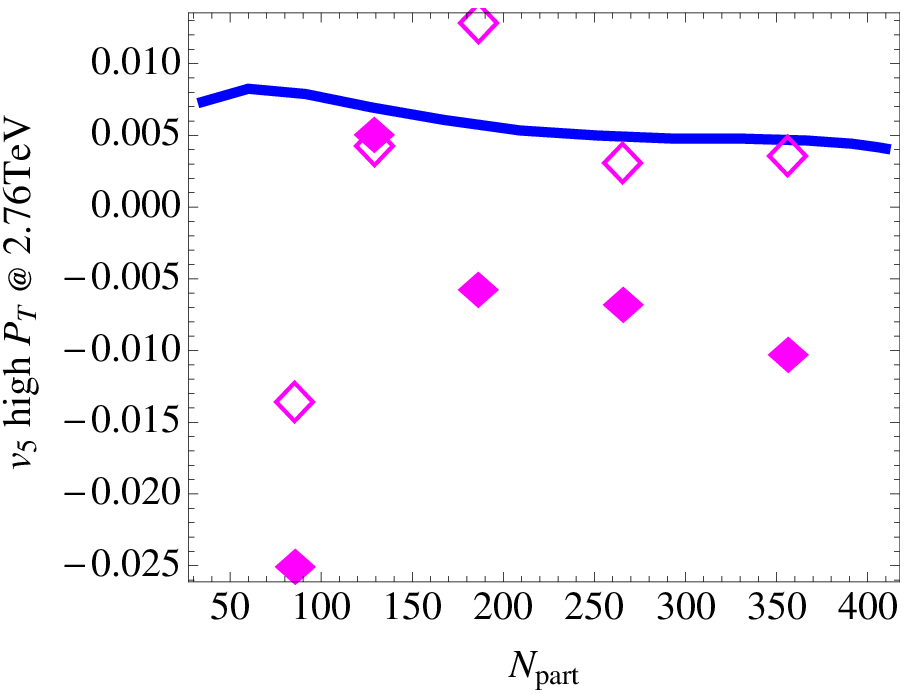}
			\includegraphics[width=8cm]{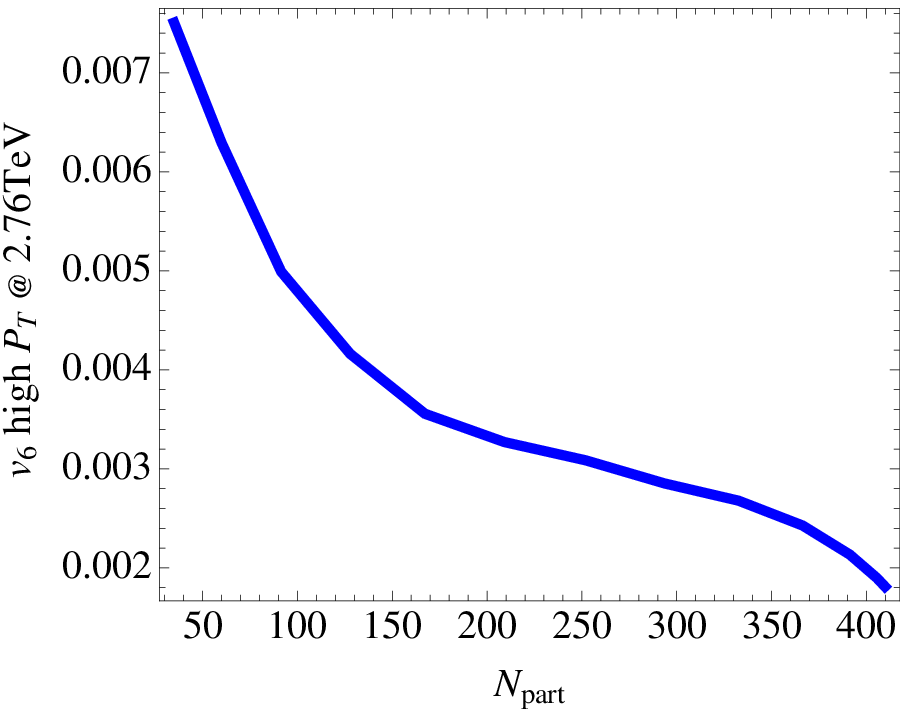}
		\caption{(color online) The $v_n$ (n=1,2,3,4,5,6) of high $p_t$ hadrons at LHC $2.76\rm\, TeV$ collisions computed from the NTcE model. In the $v_2$ plot, the data are the same as those shown in Fig.~\ref{V2_comparison}. Available data for $v_{3,4,5}$ are shown as triangles (ALICE $10-20\rm \, GeV$), open and filled diamonds (ATLAS $8-12$ and $12-20\rm \, GeV$). For the ALICE data, the $v_3$ are measured by event-plane (EP) method; the $v_4$ are measured by EP method with respect to both 2nd-harmonic EP (the filled triangles) and 4th-harmonic EP (the open triangles).  The ATLAS data are all measured by EP method. }
		\label{Vn_276}
	\end{center}
\end{figure*}

\begin{figure}[ntp]
	\begin{center}
		\includegraphics[width=10cm,height=9cm]{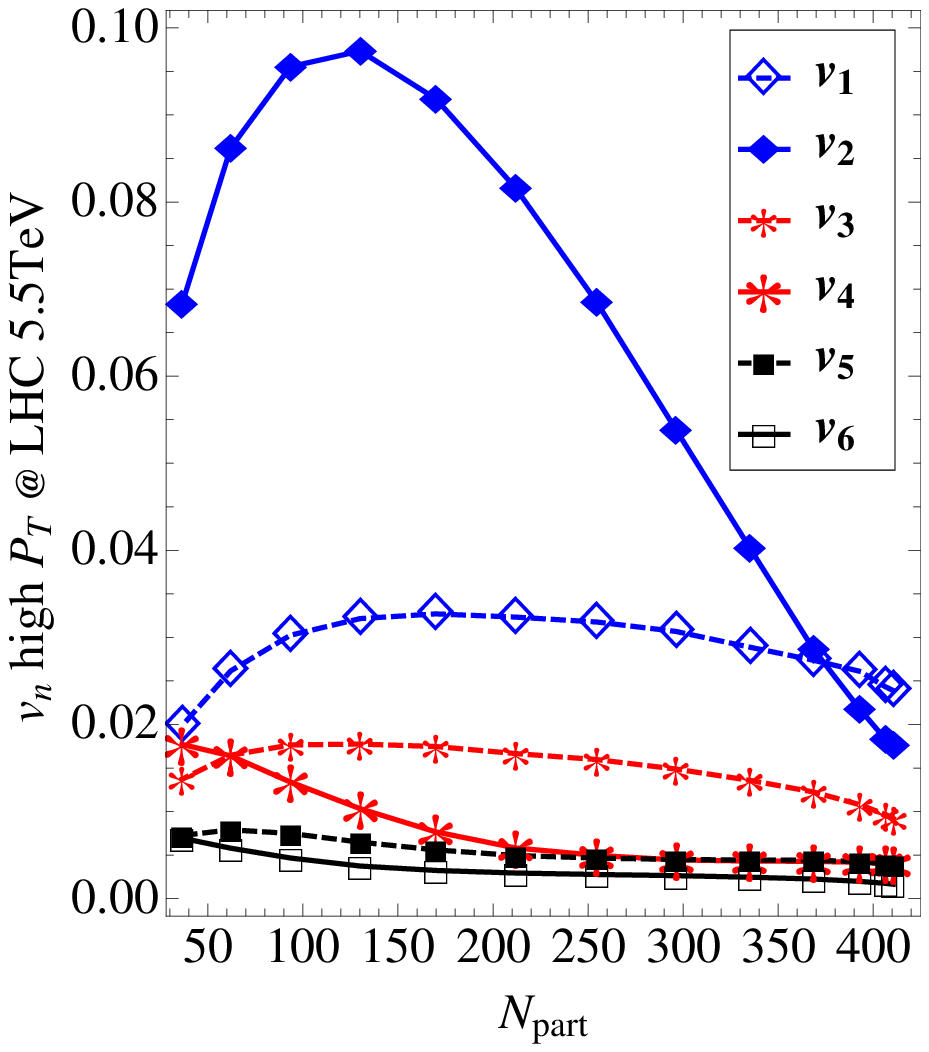}
		\caption{(color online) Predictions from NTcE model for $v_n$ (n=1,2,3,4,5,6) of high $p_t$ hadrons    at LHC $5.5\rm\, TeV$ collisions.} 
		\label{Vn_550}
	\end{center}
\end{figure}

We further come to quantify other azimuthal harmonics $v_n$(n=1,2,3,4,5,6) in the jet quenching azimuthal anisotropy based on the NTcE model. The results for LHC $2.76\rm \, TeV$ collisions are shown in Fig.~\ref{Vn_276} together with available data for the $v_{3,4,5}$\cite{Abelev:2012di,ATLAS:2012at}. 
For the ALICE data~\cite{Abelev:2012di} shown here, the $v_3$ are measured by event-plane (EP) method and the $v_4$ are measured by EP method with respect to both 2nd-harmonic EP (the filled triangles) and 4th-harmonic EP (the open triangles).  The ATLAS data~\cite{ATLAS:2012at} are measured by EP method. 
One shall however be cautious about comparing these higher harmonics with data. 
Different from $v_2$ for which both the final bulk matter event plane and the final hard response (quenching) plane are tightly correlated with the initial participant plane, for the other harmonics strongly affected by the initial state fluctuations their event plane and quenching plane are much less correlated with the initial $\epsilon_n$ plane~\cite{Jia:2012ez,HeinzMoreland2011}. 
An appropriate comparison requires an integrated event-by-event hydrodynamics plus jet quenching calculation as in \cite{Renk:2006sx,Renk:2011aa,Renk:2011qi} thus allowing a simultaneous determination of both the soft and hard anisotropies in the same event. 
Finally in Fig.~\ref{Vn_550} we show the first predictions for high-$p_t$ $v_n$(n=1,2,3,4,5,6) at LHC $5.5\rm \, TeV$ collisions, which will be tested in a few years.

\section{Summary and Discussions}

In summary we've studied event by event the hard probe of the geometry and fluctuations in the initial condition of heavy ion collisions. With  precise data sets for high $p_t$ azimuthal anisotropy at RHIC plus LHC,  jet quenching models with varied path-length dependence and matter-density dependence can be discriminated. We've found that too strong a path-length dependence (e.g. cubic) is ruled out by data at LHC,  while the model with   strong near-$T_c$ enhancement of jet-medium interaction  fully describes data at both collision energies. In addition, a full quantification of the azimuthal harmonics $v_n$ (n=1,2,3,4,5,6) of high $p_t$ hadrons is presented for LHC $2.76\rm \, TeV$ as well as $5.5\rm \, TeV$ collisions. While many other factors in modeling jet energy loss \cite{Renk:2011aa,Renk:2006sx,Renk:2011qi} also influence the jet azimuthal anisotropy and its evolution with collision energy, this study demonstrates these two's  particular sensitivity to the path-length and matter-density dependences of jet energy loss.  
Further improvements (e.g. including energy dependence, integration with realistic hydrodynamic modeling, incorporating the NTcE component with other jet quenching schemes, etc) are underway and will be reported in the future.  

It is worth emphasizing again that the near-$T_c$ enhancement model implies a strong decrease of jet-medium interaction at hotter temperature; it expects a less color-opaque medium at LHC despite only modest temperature increase  from RHIC \cite{Liao:2011kr,Zhang:2012ha}.  Consistent messages have been reported recently from a variety of independent jet quenching studies \cite{Horowitz:2011gd,Lacey:2012bg}.  The underlying picture of emergent magnetic plasma near  $T_c$  implies a more rapid running of coupling than what would be expected from perturbative picture as demonstrated in \cite{Liao:2006ry}; this seems to be consistent with the noticeable reduction of jet-medium coupling \cite{Zhang:2012ha,Betz:2012fy,Horowitz:2011gd} resulting from a rather mild $\sim 30\%$ increase of the initial fireball temperature. Therefore, in LHC's top energy heavy ion runs the created QGP might be considerably closer to the long expected asymptotically free matter. 

\acknowledgments

We thank U. Heinz, Z. Qiu, M. Gyulassy, G. Torrieri, B. Betz, A. Buzzatti, J. Jia, R. Lacey, F. Wang, D. Molnar, R. Fries, R. Rodriguez and J. Casalderrey-Solana for discussions. We also thank the RIKEN BNL Research Center for partial support.



\begin{thebibliography}{99} \frenchspacing

\bibitem{Gross:1973id} 
  D.~J.~Gross and F.~Wilczek,
  Phys.\ Rev.\ Lett.\  {\bf 30}, 1343 (1973).
  H.~D.~Politzer,
  Phys.\ Rev.\ Lett.\  {\bf 30}, 1346 (1973).

\bibitem{Collins:1974ky} 
  J.~Collins and M.~Perry,
  Phys.\ Rev.\ Lett.\  {\bf 34}, 1353 (1975).

\bibitem{Lee:1974kn} 
  T.~D.~Lee,
  Rev.\ Mod.\ Phys.\  {\bf 47}, 267 (1975).

\bibitem{Shuryak:1978ij} 
  E.~V.~Shuryak,
  Phys.\ Lett.\ B {\bf 78}, 150 (1978)
  [Sov.\ J.\ Nucl.\ Phys.\  {\bf 28}, 408 (1978)]
  [Yad.\ Fiz.\  {\bf 28}, 796 (1978)];
  J.~I.~Kapusta,
  Nucl.\ Phys.\ B {\bf 148}, 461 (1979);
  L.~D.~McLerran and B.~Svetitsky,
  Phys.\ Rev.\ D {\bf 24}, 450 (1981).

\bibitem{rhic_white_paper}
  J.~Adams {\it et al.},
  Nucl.\ Phys.\  A {\bf 757}, 102 (2005).
  K.~Adcox {\it et al.},
  Nucl.\ Phys.\  A {\bf 757}, 184 (2005).

\bibitem{Muller:2012zq} 
  B.~Muller, J.~Schukraft and B.~Wyslouch,
  Ann.\ Rev.\ Nucl.\ Part.\ Sci.\  {\bf 62}, 361 (2012)
  [arXiv:1202.3233 [hep-ex]].

\bibitem{Gyulassy:2004zy} 
  M.~Gyulassy and L.~McLerran,
  Nucl.\ Phys.\ A {\bf 750}, 30 (2005).
  E.~V.~Shuryak,
  Nucl.\ Phys.\ A {\bf 750}, 64 (2005); 
  Prog.\ Part.\ Nucl.\ Phys.\  {\bf 62}, 48 (2009).

\bibitem{Liao:2006ry}
  J.~Liao and E.~Shuryak,
  Phys.\ Rev.\  C {\bf 75}, 054907 (2007);
    Phys.\ Rev.\ Lett.\  {\bf 101}, 162302 (2008); 
  Phys.\ Rev.\ Lett.\  {\bf 109}, 152001 (2012)
  [arXiv:1206.3989 [hep-ph]];    
  M.~N.~Chernodub and V.~I.~Zakharov,
  Phys.\ Rev.\ Lett.\  {\bf 98}, 082002 (2007).
  A.~D'Alessandro and M.~D'Elia,
  Nucl.\ Phys.\ B {\bf 799}, 241 (2008).
  C.~Ratti and E.~Shuryak,
  Phys.\ Rev.\ D {\bf 80}, 034004 (2009).

\bibitem{Wang:1991xy} 
  X.~-N.~Wang and M.~Gyulassy,
  Phys.\ Rev.\ Lett.\  {\bf 68}, 1480 (1992). 
 M.~Gyulassy and M.~Plumer,
  Phys.\ Lett.\ B {\bf 243}, 432 (1990). 
 J.~P.~Blaizot and L.~D.~McLerran,
  Phys.\ Rev.\ D {\bf 34}, 2739 (1986).


\bibitem{Gyulassy:2003mc}
M.~Gyulassy, I.~Vitev, X.~N.~Wang and B.~W.~Zhang,
  arXiv:nucl-th/0302077. 
 P.~Jacobs and X.~-N.~Wang,
  Prog.\ Part.\ Nucl.\ Phys.\  {\bf 54}, 443 (2005). 
 R.~J.~Fries and C.~Nonaka,
  Prog.\ Part.\ Nucl.\ Phys.\  {\bf 66}, 607 (2011).
  J.~Casalderrey-Solana, H.~Liu, D.~Mateos, K.~Rajagopal and U.~A.~Wiedemann,  
  arXiv:1101.0618 [hep-th]. 

\bibitem{LHC_suppression} 
  K.~Aamodt {\it et al.}  [ALICE Collaboration],
  Phys.\ Lett.\ B {\bf 696}, 30 (2011). 
  
\bibitem{CMS:2012aa} 
  S.~Chatrchyan {\it et al.}  [CMS Collaboration],
  Eur.\ Phys.\ J.\ C {\bf 72}, 1945 (2012). 

\bibitem{Gyulassy:2000gk}
  M.~Gyulassy, I.~Vitev and X.~N.~Wang,
  Phys.\ Rev.\ Lett.\  {\bf 86}, 2537 (2001). 

 \bibitem{Wang:2000fq}
  X.~N.~Wang,
  Phys.\ Rev.\  C {\bf 63}, 054902 (2001).


\bibitem{Shuryak:2001me}
  E.~V.~Shuryak,
  Phys.\ Rev.\  {\bf C66}, 027902 (2002).
  A.~Drees, H.~Feng, J.~Jia,
  Phys.\ Rev.\  {\bf C71}, 034909 (2005).


\bibitem{Liao:2008dk}
  J.~Liao, E.~Shuryak,
  Phys.\ Rev.\ Lett.\  {\bf 102}, 202302 (2009).



  \bibitem{Jia:2010ee}
  J.~Jia, R.~Wei,
  Phys.\ Rev.\  {\bf C82}, 024902 (2010).

\bibitem{Jia:2011pi}
  J.~Jia, W.~A.~Horowitz and J.~Liao,
  Phys.\ Rev.\ C {\bf 84}, 034904 (2011). 


\bibitem{Betz:2011tu}
  B.~Betz, M.~Gyulassy and G.~Torrieri,
  Phys.\ Rev.\ C {\bf 84}, 024913 (2011); 
  J.\ Phys.\ G {\bf 38}, 124153 (2011)
  [arXiv:1106.4564 [nucl-th]].

\bibitem{Liao:2011kr}
  J.~Liao,
  AIP Conf.\ Proc.\  {\bf 1441}, 874 (2012)
  [arXiv:1109.0271 [nucl-th]]. 

\bibitem{Bass:2008rv} 
  S.~A.~Bass, C.~Gale, A.~Majumder, C.~Nonaka, G.~-Y.~Qin, T.~Renk and J.~Ruppert,
  Phys.\ Rev.\ C {\bf 79}, 024901 (2009). 

\bibitem{Rodriguez:2010di} 
  R.~Rodriguez, R.~J.~Fries and E.~Ramirez,
  Phys.\ Lett.\ B {\bf 693}, 108 (2010). 

\bibitem{Renk:2011aa} 
  T.~Renk,
  Phys.\ Rev.\ C {\bf 85}, 044903 (2012).

\bibitem{Zhang:2012mi} 
  X.~Zhang and J.~Liao,
  Phys.\ Lett.\ B {\bf 713}, 35 (2012). 
  
\bibitem{Zhang:2012ha} 
  X.~Zhang and J.~Liao,
  Phys.\  Rev.\  C 87, {\bf 044910} (2013)
  [arXiv:1210.1245 [nucl-th]].

\bibitem{Jia:2012ez} 
  J.~Jia,
  arXiv:1203.3265 [nucl-th].


  
\bibitem{Adler:2006bw}
  S.~S.~Adler {\it et al.} [ PHENIX Collaboration ],
  Phys.\ Rev.\  {\bf C76}, 034904 (2007).
  R.~Wei [PHENIX Collaboration],
  Nucl.\ Phys.\ A {\bf 830}, 175C (2009). 


\bibitem{Baier:1996sk}
  R.~Baier, Y.~L.~Dokshitzer, A.~H.~Mueller, S.~Peigne and D.~Schiff,
  Nucl.\ Phys.\  B {\bf 484}, 265 (1997). 
  B.~G.~Zakharov,
  JETP Lett.\  {\bf 63}, 952 (1996). 

\bibitem{Renk:2010qx} 
  T.~Renk, H.~Holopainen, U.~Heinz and C.~Shen,
  Phys.\ Rev.\ C {\bf 83}, 014910 (2011). 
  S.~Francesco, M.~Di Toro, V.~Greco, M.~Di Toro and V.~Greco,
  arXiv:1009.1261 [nucl-th].

%
\bibitem{Betz:2012fy} 
  B.~Betz and M.~Gyulassy,
  Nucl.\ Phys.\ A904-905 {\bf 2013}, 717c (2013). 
  
\bibitem{Buzzatti:2012dy} 
  A.~Buzzatti and M.~Gyulassy,
  Nucl.\ Phys.\ A904-905 {\bf 2013}, 779c (2013). 



\bibitem{Adare:2008qa} 
  A.~Adare {\it et al.}  [PHENIX Collaboration],
  Phys.\ Rev.\ Lett.\  {\bf 101}, 232301 (2008)
  [arXiv:0801.4020 [nucl-ex]].
  
\bibitem{Abelev:2012hxa} 
  B.~Abelev {\it et al.}  [ALICE Collaboration],
  Phys.\ Lett.\ B {\bf 720}, 52 (2013)
  [arXiv:1208.2711 [hep-ex]].



\bibitem{Marquet:2009eq} 
  C.~Marquet and T.~Renk,
  Phys.\ Lett.\ B {\bf 685}, 270 (2010). 

\bibitem{Adare:2010sp} 
  A.~Adare {\it et al.}  [PHENIX Collaboration],
  Phys.\ Rev.\ Lett.\  {\bf 105}, 142301 (2010). 

  
\bibitem{Renk:2006pk} 
  T.~Renk and K.~Eskola,
  Phys.\ Rev.\ C {\bf 75}, 054910 (2007).
  
  
\bibitem{Francois}
   Fran\c{c}ois~Arleo, David d'Enterria, and Andre S. Yoon, 
   	JHEP {\bf 06} 035 (2010).  

\bibitem{Bulk_harmonics}
  B.~Alver and G.~Roland,
  Phys.\ Rev.\ C {\bf 81}, 054905 (2010). 
  Z.~Qiu and U.~W.~Heinz,
  Phys.\ Rev.\ C {\bf 84}, 024911 (2011). 
  B.~Schenke, P.~Tribedy and R.~Venugopalan,
  Phys.\ Rev.\ Lett.\  {\bf 108}, 252301 (2012). 


\bibitem{miller2007}
   M.\ L.\ Miller, K.\ Reygers, S.\ J.\ Sanders, and P.\ Steinberg, 
   Annu.\ Rev.\ Nucl.\ Part.\ Sci. {\bf 57} 205 (2007).  

\bibitem{HeinzMoreland2011}
    U.~Heinz and J.~S.~Moreland, 
  Phys.\ Rev.\ C {\bf 84} 054905 (2011).  
Z.~Qiu and U.~Heinz, private communications. 

\bibitem{Hirano2009}
  Tetsufumi Hirano and Yasuishi Nara, 
  Phys.\ Rev.\ C {\bf 79} 064904 (2009).  
  Tetsufumi Hirano, Pasi Huovinen, and Yasushi Nara, 
  Phys.\ Rev.\ C {\bf 83} 021902 (2011).
  
\bibitem{Blaizot:2011xf} 
  J.~-P.~Blaizot, F.~Gelis, J.~-F.~Liao, L.~McLerran and R.~Venugopalan,
  Nucl.\ Phys.\ A {\bf 873}, 68 (2012). 


\bibitem{Renk:2006sx} 
  T.~Renk, J.~Ruppert, C.~Nonaka and S.~A.~Bass,
  Phys.\ Rev.\ C {\bf 75}, 031902 (2007). 

\bibitem{Renk:2011qi} 
  T.~Renk, H.~Holopainen, J.~Auvinen and K.~J.~Eskola,
  Phys.\ Rev.\ C {\bf 85}, 044915 (2012). 
  
\bibitem{Abelev:2012di} 
  B.~Abelev {\it et al.}  [ALICE Collaboration],
  Phys.\ Lett.\ B {\bf 719}, 18 (2013)
  [arXiv:1205.5761 [nucl-ex]].


\bibitem{ATLAS:2012at} 
  G.~Aad {\it et al.}  [ATLAS Collaboration],
  Phys.\ Rev.\ C {\bf 86}, 014907 (2012); 
  Phys.\ Lett.\ B {\bf 707}, 330 (2012). 

\bibitem{Chatrchyan:2012xq} 
  S.~Chatrchyan {\it et al.}  [CMS Collaboration],
  arXiv:1204.1850 [nucl-ex].

\bibitem{Horowitz:2011gd} 
  W.~A.~Horowitz and M.~Gyulassy,
  Nucl.\ Phys.\ A {\bf 872}, 265 (2011). 
  B.~Betz and M.~Gyulassy,
  Phys.\ Rev.\ C {\bf 86}, 024903 (2012). 
  B.~G.~Zakharov,
  JETP Lett.\  {\bf 93}, 683 (2011). 

\bibitem{Lacey:2012bg} 
  R.~A.~Lacey, N.~N.~Ajitanand, J.~M.~Alexander, J.~Jia and A.~Taranenko,
  arXiv:1202.5537 [nucl-ex]; 
  arXiv:1203.3605 [nucl-ex].

  
\end{thebibliography}
\end{document}